\documentclass[aps,pra,twocolumn,halfspacing,10pt,nofootinbib]{revtex4-1}
\usepackage{epsfig}
\usepackage{dcolumn}
\usepackage{natbib}
\usepackage[colorlinks]{hyperref}
\usepackage[utf8]{inputenc}
\usepackage[english]{babel}
\usepackage{graphicx}
\usepackage{color}
\usepackage{bm}
\usepackage{amssymb}
\usepackage[intlimits]{amsmath}
\usepackage{amsmath}
\usepackage[para,online,flushleft]{threeparttable}

\begin{document}
\title{Electric dipole moments  of atoms and molecules produced by enhanced nuclear Schiff moments}
\author{V. V. Flambaum$^{1,2}$}
\author{V.A. Dzuba$^1$}
\affiliation{$^1$School of Physics, University of New South Wales,
Sydney 2052, Australia}
\affiliation{$^2$Helmholtz Institute Mainz, Johannes Gutenberg University, 55099 Mainz, Germany}

\begin{abstract}

We perform calculations of the CP-violating atomic and molecular electric dipole moments (EDM)  induced by the interaction of the nuclear Schiff moments with electrons. EDM of atoms  Eu, Dy, Gd, Ac, Th, Pa, U, Np and Pu  are of special interest since they have isotopes with strongly enhanced nuclear Schiff moments caused by the octupole nuclear deformation or soft octupole vibration mode. These  atoms have open $4f$ or $5f$ shells making the calculations complicated. We use our special version of the configuration interaction method combined with the many-body perturbation theory method adopted for open $f$-shell case. To validate the method we perform similar calculations for simpler atoms (Xe, Hg, Tl$^+$, Pb, Pb$^{++}$, Rn, Ra) where some earlier results are available. In addition we present the estimates of the CP-violating nuclear spin - molecular axis interaction constants for molecules which may be of experimental interest including AcF, AcN, AcO$^+$, EuN, EuO$^+$, ThO, PbO and TlF. We also present updated values of the nuclear Schiff moments and atomic and molecular EDM expressed in terms of the CP-violating $\pi$-meson - nucleon interaction constants ${\bar g}_0$, ${\bar g}_1$, ${\bar g}_2$, QCD parameter ${\bar \theta}$ and quark chromo-EDMs. The results may be used to test CP-violation theories and search for axion dark matter  in atomic, molecular and solid state experiments.

\end{abstract}
\date{\today}
\maketitle

\section{Introduction}

\subsection{Nuclear Schiff moments produced by T,P-odd nuclear forces and atomic and molecular EDM produced by electric field of Schiff moments}

Measurements  of  atomic and molecular time reversal (T) and parity (P) violating electric dipole moments are used to test unification theories predicting CP-violation. They have already excluded a number  of models and significantly reduced parametric space of other popular  models  including supersymmetry \cite{PR,ERK}. Another motivation is related to the baryogenesis problem, the matter-antimatter asymmetry in the universe which is produced by an unknown CP-violating interaction. The expected magnitude of an  EDM is  very small, therefore, we are looking for mechanisms that enhance the effects - see e.g.  \cite{Khriplovich,KL,GF}. 

Schiff demonstrated that  the nuclear EDM is completely screened  in neutral atoms and molecules  and noted that a nonzero atomic EDM still may be produced if the distribution of EDM and charge in a nucleus are not proportional to each other \cite{Schiff}.  Further works  \cite{Sandars,Hinds,SFK,FKS1985,FKS1986} introduced and calculated the so called Schiff moment, a vector moment presenting electric field inside the nucleus after taking into account nuclear EDM screening by electrons. 
This Schiff moment electric field polarizes the  atom and produces  an  atomic EDM directed along the nuclear spin. Refs. \cite{Sandars,Hinds} calculated  the Schiff moment due to the  proton EDM.  Refs. \cite{SFK,FKS1985,FKS1986} calculated (and named) the nuclear Schiff moment  produced by the P,T-odd nuclear forces. 
It was shown in \cite{SFK} that the contribution of the P,T-odd forces to the nuclear EDM and Schiff moment is  larger than the contribution of a nucleon EDM.   In Ref. \cite{FG} an accurate expression for the Schiff moment electrostatic potential has been derived and the finite nuclear size corrections to the Schiff moment operator introduced (see also \cite{AKozlovC,AKozlovA}).     

\subsection{Enhancement of Schiff moment due to nuclear octupole deformation and soft octupole vibration mode}

A number  of nuclei have an opposite parity level with the same spin close to the ground state. This may lead to an enhancement  of the nuclear Schiff moment produced by the P,T-odd nuclear forces which admix this close state wave function to the ground state \cite{SFK} \footnote{Nuclear EDM and magnetic quadrupole produced by the T,P-odd nuclear forces are also enhanced  due to an opposite parity level with the same spin close to the ground state \cite{HH,SFK}. Collective enhancement of the magnetic quadrupole moments in deformed nuclei have been demonstrated in \cite{F1994}.} .  
However, the largest enhancement ($\sim 10^2 - 10^3$ times) happens in nuclei with an intrinsic octupole deformation where both, the 
small energy difference of nuclear levels with opposite parity and the collective effect work together \cite{Auerbach,Spevak}. According to  \cite{Auerbach,Spevak} this happens in some isotopes of Fr, Rn, Ra and actinide atoms. Atomic and molecular EDMs produced by the Schiff moment 
increase with the nuclear charge $Z$ faster than $Z^2$  \cite{SFK}. This is another reason why EDM in actinide atoms and their molecules are expected to be significantly larger than in other systems. 

The Schiff moment is proportional to the squared octupole deformation parameter $(\beta_3)^2$ which is about $(0.1)^2$ \cite{Spevak}. According to Ref. \cite{Engel2000}, in nuclei with a soft octupole vibration mode the squared dynamical octupole deformation  $<(\beta_3)^2> \sim (0.1)^2$, i.e. it is the same as the static octupole deformation. This means that a similar enhancement of the Schiff moment may be due to the dynamical octupole effect \cite{Engel2000,FZ,soft2} in nuclei where 
 $<\beta_3>=0$ \footnote{Recall an ordinary oscillator where $<x>=0$ while $<x^2>$ is not equal to zero.}. This observation significantly increases the list of nuclei where the Schiff moment is enhanced. 
 
 In the papers \cite{Auerbach,Spevak,EngelRa,Jacek2018}   numerical calculations of  Schiff moments and estimates of atomic EDM produced by the electrostatic interaction between electrons and these moments have been done  for  $^{223}$Ra, $^{225}$Ra, $^{223}$Rn, $^{221}$Fr, $^{223}$Fr, $^{225}$Ac and $^{229}$Pa. 
 Unfortunately, these nuclei have a short lifetime. Several experimental groups have considered experiments with  $^{225}$Ra and $^{223}$Rn       \cite{RaEDM,RaEDM2,RnEDM}. 
    The only published EDM measurements  \cite{RaEDM,RaEDM2}  have been done for $^{225}$Ra which has 15 days half-life. In spite of the Schiff moment enhancement the $^{225}$Ra  EDM measurement has not reached yet the sensitivity to the T,P-odd interaction  comparable to the Hg EDM experiment \cite{HgEDM}. The experiments continue, however,  the instability of $^{225}$Ra and a relatively small number of atoms available  may be a problem. In Ref. \cite{Th} the    
nuclear Schiff moment of $^{229}$Th nucleus has been estimated since this nucleus has a much longer lifetime (7917 years).  In the Ref.\cite{FF19} the list of the candidates for the enhanced Schiff moments has been extended to include
stable isotopes  $^{153}$Eu, $^{161}$Dy, $^{163}$Dy, $^{155}$Gd, and long lifetime nuclei   $^{235}$U, $^{237}$Np, $^{233}$U, $^{229}$Th, $^{153}$Sm, $^{165}$Er, $^{225}$Ac, $^{227}$Ac, $^{231}$Pa, $^{239}$Pu. The estimates of the enhanced Schiff moments have been done for the most attractive cases of  $^{153}$Eu, $^{235}$U, $^{237}$Np and $^{227}$Ac. 

In this paper we  present updated values for many nuclear Schiff moments  expressed in terms of the CP-violating $\pi$-meson - nucleon interaction constants ${\bar g}_0$, ${\bar g}_1$, ${\bar g}_2$, QCD parameter ${\bar \theta}$ and quark chromo-EDMs. We also calculate  atomic and molecular EDMs induced by these Schiff moments.

\subsection{Oscillating Schiff moments and atomic and molecular electric dipole moments produced by axion dark matter}

The CP-violating  neutron EDM may be due to the QCD $\theta$-term \cite{Witten}.  It was noted in Ref.  \cite{Graham} that the  axion dark matter produces oscillating neutron EDM since the axion field  is equivalent to the oscillating ${\bar \theta}$.  QCD $\theta$-term also produces P,T-odd nuclear forces creating nuclear Schiff moments. Correspondingly, axion field  also produces oscillating nuclear Schiff moments \cite{Stadnik} which are  enhanced  by the octupole mechanism. To obtain the results for the oscillating Schiff moment it is sufficient to replace the constant ${\bar \theta}$ by ${\bar \theta}(t)= a(t)/f_a$, where $f_a$ is the axion decay constant, $a(t) =a_0 \cos{m_a t}$, $(a_0)^2= 2 \rho /(m_a)^2$, $\rho$ is the axion dark matter energy  density \cite{Graham,Stadnik}.  Moreover, in the case of the resonance between the frequency of the axion field oscillations and molecular transition frequency there may be an enormous  resonance enhancement of the oscillating nuclear Schiff moment  effect  \cite{OscillatingEDM}. Since an oscillating nuclear Schiff moment and oscillating nuclear EDM may be produced by the axion dark matter, corresponding measurements may be used to search for the dark matter. First results of such search have been published in Ref. \cite{nEDM}, where the oscillating neutron EDM and oscillating $^{199}$Hg Schiff moment have been measured.  Search for the effects produced by the  oscillating axion-induced Schiff moments in solid state materials is in progress \cite{Casper}. 

\section{Estimates of  nuclear Schiff moments} \label{Schiff}
In this section we  present values of the nuclear Schiff moments for all atoms considered in the present work. We present corrected values if there is  a reason to make the corrections, express the results in terms of the $\pi$-meson - nucleon interaction constants   ${\bar g}_0$, ${\bar g}_1$, ${\bar g}_2$, QCD parameter ${\bar \theta}$ and quark chromo-EDMs (such presentations were often not available) and perform rough estimates for several nuclei where the Schiff moments were unknown.

\subsection{Estimates of enhanced Schiff moments  in deformed nuclei}
 
Nuclear spectra of a nucleus with the octupole (pear-shape) deformation are similar to  spectra of a diatomic molecule made from different atoms \footnote{Note that the enhancement of the EDM and Schiff moment in nuclei with the octupole deformation is similar to the enhancement of the T,P-violating effects in polar molecules with non-zero electron angular momentum which have doublets of the opposite parity levels \cite{SushkovFlambaum}.}. Therefore, looking into the spectra gives us  first indication of the octupole\footnote{The doublet splitting in molecules is due to the Coriolis interaction. In nuclei the splitting is dominated by the
"tunnelling" of the octupole bump to other side of the nucleus causing change of the valence nucleon spin projection
to the nuclear axis. In fact, it is just an octuple vibration mode, so there is no sharp  boundary between the static
 deformation in the minimum of the potential energy  and soft octuple vibration when this minimum is very shallow
or does  not exist. Note that contrary to the Coriolis splitting in diatomic molecules the doublet splitting due the tunnelling does not increase with the rotational angular momentum - see the nuclear spectra in Ref.  \cite{nndt}.}. Other indications include measured probabilities of the electric octupole and electric dipole transitions. There are also  sophisticated nuclear calculations which give us calculated deformation parameters - see e.g. \cite{Afanasjev2016,Robledo,Robledo2013,Butler,Sm152,Minkov,Nomura2015,Bernard2016,Fu2018,Bucher2017}.
 
Schiff moment is defined by the following expression \cite{SFK}: 
\begin{equation}\label{S}
{\bf S}=\frac{e}{10} [<r^2 {\bf r}> - \frac{5}{3Z}<r^2><{\bf r}>], 
\end{equation}
where $<r^n> \equiv \int \rho({\bf r}) r^n d^3r$ are the moments of the nuclear charge density $\rho$. The second term originates from the electron screening and contains  nuclear mean squared charge radius  $<r^2>/Z$ and nuclear EDM  $d=e<{\bf r}>$, where $Z$ is the nuclear charge. 

If a nucleus has an octupole deformation $\beta_3$ and a quadrupole deformation $\beta_2$, in the fixed-body  (rotating) frame the Schiff moment $S_{intr}$ is proportional to the octupole moment $O_{intr}$, i.e. it  has a collective nature  \cite{Auerbach,Spevak}:
\begin{equation}\label{Sintr}
 S_{intr} \approx \frac{3}{5 \sqrt{35}} O_{intr} \beta_2 \approx    \frac{3}{20 \pi  \sqrt{35}} e Z R^3 \beta_2 \beta_3 ,
\end{equation}
where $R$ is the nuclear radius.
 However, in the laboratory frame EDM and  Schiff moment are forbidden by the parity and   time reversal invariance. Indeed, EDM and Schiff moment are polar $T$-even vectors which must be directed along the nuclear spin $I$ which is $T$-odd pseudovector.  
   
   Nucleus with an octupole deformation and non-zero nucleon angular momentum  has a doublet of close opposite parity rotational states $|I^{\pm}>$ with the same angular momentum $I$ ($| I^{\pm} >=\frac{1}{\sqrt{2}} (|\Omega> \pm |-\Omega>)$, where $\Omega$ is  the projection of $I$ on to the nuclear axis). The states of of this doublet are mixed by $P,T-$violating interaction $W$. The mixing coefficient is:
\begin{equation}\label{alpha}
 \alpha_{+-}=\frac{<I^-| W| I^+>}{E_+  -  E_-} . 
\end{equation}
This mixing polarises  nuclear axis ${\bf n}$ along the nuclear spin ${\bf I}$,  $<n_z>= 2 \alpha _{+-}\frac{I_z}{I+1}$,
and the intrinsic Schiff moment shows up in the laboratory frame \cite{Auerbach,Spevak}:
\begin{equation}\label{Scol}
 S= 2 \alpha_{+-} \frac{I}{I+1} S_{intr}. 
\end{equation}
   According to Ref. \cite{Spevak} the T,P-violating matrix element is approximately equal to
   \begin{equation}\label{W}
   <I^-| W| I^+> \approx \frac{\beta_3 \eta}{A^{1/3}} [ \textrm{eV}].
  \end{equation}  
  Here $\eta$ is the dimensionless strength constant of the nuclear $T,P$- violating potential $W$:
   \begin{equation}\label{eta}
 W= \frac{G}{\sqrt{2}} \frac{\eta}{2m} ({\bf \sigma \nabla}) \rho ,
   \end{equation}
where $G$ is the Fermi constant, $m$ is the  nucleon mass and $\rho$ is the nuclear number density. Eqs. (\ref{Sintr},\ref{alpha},\ref{Scol},\ref{W}) give analytical estimate for the  Schiff moment \cite{Spevak,FF19}: 
\begin{equation}\label{San}
 S \approx 1. \cdot 10^{-4} \frac{I}{I+1} \beta_2 (\beta_3)^2 Z A^{2/3} \frac{[\textrm{KeV]}}{E_-  -  E_+} e \,\eta \, [\textrm{fm}^3],
   \end{equation}
This estimate is in agreement with more accurate numerical calculations available for  a number of nuclei \cite{Spevak}.
 For example, it gives $S=280 \, e \,\eta \, \textrm{fm}^3$ for $^{225}$Ra, which practically coincides with the result  of the numerical calculation in Ref.  \cite{Spevak} $S=300 \, e \,\eta \, \textrm{fm}^3$. 

 Within the meson exchange theory, the $\pi$-meson exchange gives the dominating contribution to the T,P-violating nuclear forces \cite{SFK}. In the standard notations $g$ is the strong $\pi$-meson - nucleon  interaction constant and ${\bar g}_0$, ${\bar g}_1$, ${\bar g}_2$ are the  $\pi$-meson - nucleon CP-violating interaction  constants in the isotopic channels $T=0,1,2$  \footnote{We also estimated contribution of the exchange by $\eta$-meson  which is 4 times heavier than $\pi$-meson and  is usually  assumed to give a  smaller contribution. Indeed, the second power of the meson mass appears in the denominator of the effective interaction constant for the  meson-induced nucleon interaction, so the expected suppression is 1/16. However, the $\eta$-meson CP-violating exchange constant ${\bar g}$ is an order of magnitude larger than the $\pi$-meson constant $ {\bar g}_0$ \cite{Vries2015}. In addition,  the $\eta$-meson CP-violating contribution has the same sign for protons and neutrons contrary to  the   $\pi$-meson one. As a result, the suppression of the   $\eta$-meson contribution to the nuclear Schiff moment is few times only. Therefore, future more accurate calculations of the Schiff moment should include  the $\eta$-meson contribution as well as the finite nuclear size corrections found in Ref. \cite{FG}. }. 

One can express the results in terms of more fundamental parameters such as the QCD $\theta$-term constant  ${\bar \theta}$ using the relation 
$|g {\bar g}_0|=0.37|{\bar \theta}|$ from Ref. \cite{Witten} or updated results  \cite{Yamanaka2017,Vries2015,Bsaisou2015},
       $g \bar{g}_0 = 0.21 \bar{\theta} $,       $g \bar{g}_1 = -0.046 \bar{\theta}$,  which give  practically the same  value of $S(\bar{\theta})$.
Alternatively, the results can be expressed via the  quark chromo-EDMs ${\tilde d_u}$ and  ${\tilde d_d}$: $g {\bar g}_0 = 0.8 \cdot10^{15}({\tilde d_u} +{\tilde d_d})$/cm,  $g {\bar g}_1 = 4 \cdot 10^{15}({\tilde d_u} - {\tilde d_d})$/cm  \cite{PR}.

 Numerical calculations performed in Ref. \cite{EngelRa} found  the Schiff moment of  $^{225}$Ra in terms of $g {\bar g_{0,1,2}}$. We also  express  the Schiff moment of  $^{225}$Ra in terms of $\bar{\theta} $ and $ {\tilde d_u}$ and  ${\tilde d_d}$:
\begin{align}
 S(^{225}{\rm Ra},g) & \approx   (  - 2.6 g {\bar g}_0 +  12.9 g {\bar g}_1 -6.9 g {\bar g}_2)\, e\cdot \textrm{fm}^3\, , \nonumber \\
 S(^{225}{\rm Ra}, {\bar \theta})& \approx  - \,{\bar \theta} \, e\cdot \textrm{fm}^3 \ , \label{Raschiff} \\
 S(^{225}{\rm Ra},{\tilde d})& \approx  10^4 ( 0.50 \,{\tilde d}_u -  0.54 \,{\tilde  d}_d )\, e\cdot \textrm{fm}^2 \, . \nonumber
 \end{align}
The analytical formula for the Schiff moment Eq. (\ref{San})  gives us dependence of the Schiff moment on the nuclear parameters. Nucleus  $^{225}$Ra  has  the octupole  deformation
$\beta_3$=0.099, quadrupole deformation 
  $\beta_2$=0.129, nuclear spin $I=1/2$ and interval between the opposite parity levels $E(1/2^-)- E(1/2^+)$=55.2 KeV \cite{Spevak}. Using Eqs. (\ref{Raschiff},\ref{San}) we present  the result for other nuclei with the octupole deformation in the following form    \cite{Th,FF19}:
\begin{align}\label{Sg} 
  S(g) & \approx  K_S(  - 2.6 g {\bar g}_0 +  12.9 g {\bar g}_1 -6.9 g {\bar g}_2)\, e\cdot \textrm{fm}^3\, ,\\
  \label{Stheta} 
  S({\bar \theta})& \approx  - K_S\,{\bar \theta} \, e\cdot \textrm{fm}^3 \ ,\\
\label{Sd} 
 S({\tilde d})& \approx  10^4 K_S ( 0.50 \,{\tilde d}_u -  0.54 \,{\tilde  d}_d )\, e\cdot \textrm{fm}^2 \, ,
\end{align}   
where $K_S=K_I K_{\beta}K_A K_E$, $K_I=\frac{3 I}{I+1}$, $K_{\beta}=791\beta_2 (\beta_3)^2$, $K_A=0.00031 Z A^{2/3}$,  $K_E= \frac{55\textrm{KeV}}{E^-  -  E^+} $. 
By definition, numerical
factors are chosen such that  these coefficients are equal to 1 for  $^{225}$Ra and are of the order  of unity for other heavy nuclei with octupole deformation. The values of $K_S$ for deformed nuclei with strongly enhanced  collective Schiff moments  are presented in the Table \ref{t:Schiff}. Below we present the explanation how these results were obtained.

$^{227}$Ac nucleus  has a half-life of 21.8 years. It  is produced commercially for cancer treatment.  
The half-life of $^{237}$Np is 2.14 million years. It is produced in  macroscopic quantities in nuclear reactors. In Ref. Ref. \cite{FF19} we obtained $K_S=10$ for $^{227}$Ac  and $K_S=6$ for  $^{227}$Np.  In Ref. \cite{FF19} we used  calculated values $\beta_3=0.134$ for  $^{227}$Ac and   
$\beta_3=0.12$ for $^{227}$Np.  However, in the experimental paper Ref.  \cite{Ac} it was found that   $\beta_3$ is 0.07 for $^{229}$Ac and 0.1 for $^{227}$Ac and $^{225}$Ac. Updated values of the nuclear  Schiff moments for $^{227}$Ac ($K_S=6$) and $^{237}$Np ($K_S=4$) are obtained  by  multiplying the results of Ref. \cite{FF19}  by the ratio of  $(\beta_3)^2$ from  Ref.  \cite{Ac} and Ref. \cite{FF19}. 

$^{153}$Eu  is stable with 52\% natural abundance. Its nuclear  spectra indicate  octupole deformation (since they have rotational doublets, with the same values of the moment of inertia for opposite parity states in the doublets - see details in Ref. \cite{FF19}).   According to Ref. \cite{FF19} for  $^{153}$Eu  $K_S=3.7$

 According to Ref. \cite{Spevak} the Schiff moment of $^{225}$Ac is predicted to be three times larger than that of  $^{225}$Ra, i.e. $K_S=3$ \footnote{In Ref. \cite{Spevak} we performed calculations of the Schiff moment produced by the contact T,P-violating nuclear potential (\ref{eta}). Corresponding interaction constants may be expressed in terms of the $\pi$-meson exchange constants 
 $g {\bar g}_{0,1,2}$ (see e.g. \cite{FDK}) which dominate the T,P-violating nuclear potential and are widely used now. To avoid lengthy discussion of this problem we often present the result as a ratio of the Schiff moment of a given nucleus to that of  $^{225}$Ra which has been calculated in Ref. \cite{EngelRa} in terms of $g {\bar g}_{0,1,2}$.}.
 
Nuclear spectra of $^{222}$Rn nucleus \cite{nndt} indicate  octupole deformation. The nucleus  $^{223}$Rn  has a neutron above $^{222}$Rn nucleus and probably has the octupole deformation too or at least the soft octupole vibration mode.  According to Ref. \cite{Spevak}  the Schiff moment of $^{223}$Rn exceeds the Schiff moment of 
$^{225}$Ra 3 times, i.e. $K_S=3$. 

 Nuclear spectra of $^{239}$Pu nucleus \cite{nndt} indicate octupole deformation. However, the energy interval between the levels of the doublet in $^{239}$Pu is 9 times larger than in $^{225}$Ra:   $E(1/2^-)- E(1/2^+)$= 470 KeV. The deformation parameters have been calculated in Ref. \cite{Afanasjev2016} for the even-even isotope  $^{240}$Pu:    $(\beta_3)^2  =(0.066)^2$  and $\beta_2$=0.284 \footnote{Note that different calculations of the deformation parameters  use different nucleon interaction models and may give significantly different results. Moreover, sometimes the authors come to different conclusions about the existence of the octupole deformation - see e.g.  Refs. \cite{Afanasjev2016,Robledo,Robledo2013,Butler,Sm152,Minkov,Nomura2015,Bernard2016,Fu2018,Bucher2017}. This is why we look for the signatures of the octupole deformation in the experimental nuclear excitation spectra. We use deformation parameters calculated in Ref. \cite{Afanasjev2016} since this paper contains the most comprehensive list of  $\beta_3$ values for  even-even nuclei. The values of  $\beta_3$ in different nucleon interaction models in  Ref. \cite{Afanasjev2016}  are practically the same (within the accuracy of our estimates of the Schiff moment). Therefore, we present only one number for $\beta_3$.}. 
 Using these parameters  and Eq. (\ref{Sg}) we  obtain $K_S \sim 0.12$.
Note that $^{239}$Pu may be considered as $^{238}$Pu nucleus plus neutron. The nuclear spectra of  $^{238}$Pu are consistent with the octupole deformation. Further study of this problem should bring us more reliable information about the octupole deformation parameter $\beta_3$ and  more accurate estimates of the  $^{239}$Pu Schiff moment.

According to Ref. \cite{Spevak} the Schiff moment of $^{223}$Fr is predicted to exceed the Schiff moment of $^{225}$Ra 1.6 times. This gives $K_S=1.6$.
Ref. \cite{Spevak} gives the Schiff moment of $^{221}$Fr equal to  0.14 of the  $^{225}$Ra Schiff moment, i. e. $K_S=0.14$.

The nuclear spectra of  $^{232}$U (as well as the spectra of $^{234}$U, $^{236}$U and $^{238}$U)  are consistent with  the octupole deformation. The nucleus $^{233}$U may be considered as $^{232}$U nucleus plus neutron.  Nuclear spectra of $^{233}$U nucleus \cite{nndt}  are also consistent with the octupole deformation. However, the energy interval between the levels of the doublet in $^{233}$U is 5 times larger than in $^{225}$Ra:   $E(5/2^-)- E(5/2^+)$= 299 KeV. The deformation parameters have been calculated in Ref. \cite{Afanasjev2016} for the even-even isotope  $^{232}$U:    $(\beta_3)^2  =(0.17)^2$  and $\beta_2$=0.238.  Using these parameters  and Eq. (\ref{Sg}) we  obtain $K_S \sim 2$.
 This estimate indicates a possibility of a large Schiff moment  in $^{233}$U. However, further experimental and theoretical investigation is needed.

Ref. \cite{HH} suggested a very interesting case of $^{229}$Pa which probably has a level of opposite parity and the same angular momentum very close to the ground state. The latest measurement \cite{229Pa} gave a position of this level at 60$\pm$50 eV.   According to Ref. \cite{Spevak}  the Schiff moment of $^{229}$Pa is predicted to exceed the Schiff moment of 
$^{225}$Ra 40 times. 
 We should note that the estimate $K_S=40$ is  valid if the close level 60$\pm$50 eV really exists and forms the rotational doublet  related to the octupole deformation of  $^{229}$Pa with the ground state. So far there is no truely convincing  evidence for such case.

The nucleus  $^{161}$Dy is stable. The spectrum of $^{161}$Dy is consistent with the octupole deformation, several rotational doublets are seen. The interval between the opposite parity levels is only 25 KeV,  it is 4 times smaller than in $^{153}$Eu. However, we have not found  any calculations giving the octupole deformation in $^{161}$Dy or even-even nucleus $^{160}$Dy.
Therefore, a conservative result for  the Schiff moment of $^{161}$Dy (based on the extrapolation from  $^{153}$Eu ) may be presented as an upper estimate $K_S \lesssim 4$.  
In another stable isotope   $^{163}$Dy the interval between the opposite parity levels is 10 times larger than in  $^{161}$Dy, therefore the upper estimate for the Schiff moment is an order of magnitude smaller.
 
We will try to provide some information for other nuclei  of interest.  While studying experimental nuclear spectra we  noted in Ref. \cite{FF19} a possible trend: 
adding proton  to an even-even nucleus with  octupole deformation usually supports the octupole deformation. Possibly, this is due  to proton increasing the Coulomb repulsion. 
However, adding neutron to an even-even nucleus with octupole deformation sometimes  blur  the features of the rotational spectrum for the  octupole \footnote{We must admit that a conventional point of view is  that  the influence of shell structure  dominates the effects of adding an odd particle to a core. Moreover, review \cite{Butler}  suggested that  both odd proton and odd neutron have a stabilizing effect on octupole deformation. The picture is not completely settled yet. Therefore, we do no automatically assume that odd nucleon always produces stabilising effect on the octupole. Instead, we look for the experimental nuclear spectra with two nearly parallel rational bands (i.e. corresponding to approximately the same moment of inertia) starting from the same value of nuclear spin and having opposite parity (i.e. we look for rotational bands  with several opposite parity doublets).}.
 Indeed, nuclear spectra in  $^{229}$Th, $^{235}$U, $^{153}$Sm, $^{165}$Er,   $^{155}$Gd  do not provide clear evidence for the octupole deformation but have at least one doublet of the opposite parity levels with the same nuclear spin. The probable outcome for such nuclei is existence of the soft octupole vibrational mode which produces some enhancement of the Schiff moments. 
 
  For example, $^{228}$Th has signatures of the octupole deformation in its rotational spectrum.  However, adding a neutron to this nucleus and forming    $^{229}$Th seems to blur  the features of the rotational spectrum for the  octupole.  The estimate for the $^{229}$Th Schiff moment has been done in Ref. \cite{Th}.
 According to Ref. \cite{Minkov}, $^{229}$Th nucleus has the octupole deformation with the parameters $\beta_3$=0.115,   $\beta_2$=0.240. The nuclear spin  $I=5/2$ and the  interval between the opposite parity levels is $E(5/2^-)- E(5/2^+)$=133.3 KeV.  Eq. (\ref{Sg}) gives us the value of the Schiff moment which is twice larger than that of
$^{225}$Ra. However, the static octupole deformation in $^{229}$Th does not explicitly show up in the nuclear rotational spectra. In fact,  we only see one doublet of the opposite parity states with the same spin. In this situation we may only offer an  upper limit  $K_S \lesssim  2$.

 $^{235}$U  is practically stable (half life 0.7 billion years), with 0.75\% natural abundance. The interval between opposite parity levels which are mixed by the T,P-odd interaction is $E(\frac{7}{2}^+) - E(\frac{7}{2}^-)=81.7$ KeV.  This nucleus has a neutron above  $^{234}$U nucleus which according to Ref.~\cite{Afanasjev2016} has octupole deformation with  $(\beta_3)^2  =(0.17)^2$  and $\beta_2$=0.25.  However, experimental nuclear excitation spectra of $^{235}$U  do not show  parity doublets for higher rotational states.
Assuming that there is a soft vibrational octupole mode in $^{235}$U we obtained an upper estimates $K_S \lesssim 3$  \cite{FF19}.

  Nuclei  with a valence neutron  $^{155}$Gd, $^{153}$Sm, $^{165}$Er  are close to the area of even-even nuclei with the  octupole deformation, they have opposite parity level with the same spin close to the ground state  but their spectra do not indicate octupole deformation. We may expect an order of magnitude enhancement in comparison with the Schiff moments in  nuclei which do not have close level of opposite parity such as $^{199}$Hg and  $^{207}$Pb considered below.
 
\subsection{Schiff moments in Xe, Hg, Tl and Pb nuclei} 

Let us now consider Schiff moments of nearly spherical nuclei where the Schiff moments are not enhanced. The Schiff moment of $^{199}$Hg was firstly calculated in Ref. \cite{FKS1986}. The most complete calculation in terms of $\pi$-meson interaction constants $g {\bar g}_{0,1,2}$  has been performed in Ref. \cite{HgCalc} using 5 different interaction models. We present average of these 5 values and also express the result in terms of  ${\bar \theta}$ and quark chromo-EDMs ${\tilde d_u}$ and  ${\tilde d_d}$. The nucleus $^{207}$Pb has the same spin and parity ($I^P=1/2^{-}$) and close value of the magnetic moment (0.59 nuclear magnetons)  to that of $^{199}$Hg (0.51 nuclear magnetons), i.e. the valence nucleons in  $^{207}$Pb and $^{199}$Hg occupy the same orbital. Therefore, within the theoretical  accuracy Schiff moments of $^{199}$Hg and $^{207}$Pb  are equal:
 \begin{align}\label{HgPbschiff}
 \nonumber
 S(^{207}{\rm Pb},g)  \approx  S(^{199}{\rm Hg},g)  \\
 \approx (  0.023 g {\bar g}_0 -0.007  g {\bar g}_1 +0.029 g {\bar g}_2)\, e\cdot \textrm{fm}^3 ,\\
 S(^{207}{\rm Pb},{\bar \theta})  \approx  S(^{199}{\rm Hg},{\bar \theta})  \approx 0.005 \,{\bar \theta} \, e\cdot\textrm{fm}^3 \ ,\\
 S(^{207}{\rm Pb},{\tilde d})  \approx  S(^{199}{\rm Hg},{\tilde d}) \approx 5 {\tilde d}_d\, e\cdot \textrm{fm}^2 \, .
 \end{align}
Note that these results include contributions of the T,P-odd nuclear forces between the nucleons and the nucleon electric dipole moments (as it was done  in Ref. \cite{HgCalc}), however, the contribution of the proton and neutron EDM to the nuclear Schiff moments is significantly smaller than the contribution of the T,P-odd nuclear forces (firstly this was pointed out in Ref. \cite{SFK}). In Refs. \cite{DFGK02,DS03} contributions of the neutron $d_n$ and proton $d_p$ EDM to  $S(^{199}{\rm Hg})$  have been presented separately:
$S(^{207}{\rm Pb},d)  \approx  S(^{199}{\rm Hg},d)=(1.9 d_n +0.2 d_p)\, \textrm{fm}^2$.
However, in Ref.  \cite{HgCalc} the  contribution of the neutron EDM (averaged over 5 interaction models) is 3 times smaller. Then taking the ratio of the neutron and proton contributions  from Ref. \cite{DFGK02} (where the result is based on the fitting of the neutron and proton contributions to the nuclear magnetic moment) we obtain 
\begin{equation} \label{dHg2}
S(^{207}{\rm Pb},d)  \approx  S(^{199}{\rm Hg},d)=(0.6  d_n +0.06 d_p) \, \textrm{fm}^2
\end{equation}

Schiff moments of $^{203}$Tl and $^{205}$Tl in the Saxon-Woods nuclear potential have been calculated in Ref. \cite{FKS1986} in terms of the contact nucleon-nucleon T,P-violating interaction.  Here we  express the results in terms of the more fundamental interaction constants:
\begin{align}\label{Tlschiff}
 \nonumber
 S(^{203}{\rm Tl},g)  \approx  S(^{205}{\rm Tl},g)  \\
 \approx (  0.13 g {\bar g}_0 -0.004  g {\bar g}_1 -0.27  g {\bar g}_2)\, e\cdot \textrm{fm}^3 ,\\
 S(^{203}{\rm Tl},{\bar \theta})  \approx  S(^{205}{\rm Tl},{\bar \theta})  \approx 0.027  \,{\bar \theta} \, e\cdot\textrm{fm}^3 \ ,\\
 S(^{203}{\rm Tl},{\tilde d})  \approx  S(^{205}{\rm Tl},{\tilde d}) \approx (12 {\tilde d}_d +9 {\tilde d}_u)\, e\cdot \textrm{fm}^2 \, .
 \end{align}
 In the case of $^{199}$Hg the  sophisticated Hartree-Fock-Bogoliubov calculations  in Ref. \cite{HgCalc} gave  a smaller value of the nuclear Schiff moment than the calculations  in Ref. \cite{FKS1986} performed in the Saxon-Woods potential. In  Hg valence nucleon is neutron, therefore, nuclear EDM and Schiff moment appear due to the nuclear core polarization \cite{FKS1986}. $^{203}$Tl and $^{205}$Tl nuclei have  valence proton, therefore EDM and  Schiff moment appear even without the core polarization.  Still the many-body corrections may play an important role \cite{FKS1986} and provide up to factor of 3 suppression of $S(^{203,205}{\rm Tl})$ .
 
 We may also separate the contribution of the valence proton EDM to the nuclear Schiff moment. The result from Refs. \cite{Sandars,Hinds},  $S=- d_p (R/6)$
 ,  depends on the difference $R\equiv r_v^2- r_q^2$ of the squared values of the spin distribution radius $r_v^2$ and charge radius $r_q^2$. Unfortunately, even the sign of $R$ is not firmly established  \cite{SFK,CS83,FDK}. Using the average result of the Hartree Fock calculations with different interactions performed by Alex Brown and presented in Ref. \cite{CS83},    $R = 2.3 $ fm$^2$,  we obtain  $S= 0.4 d_p $ fm$^2$.
 
 Finally, we present the Schiff moment of  $^{129}$Xe which has valence neutron. Firstly, it was calculated in the Saxon-Woods potential  in Ref. \cite{FKS1986}. Then Ref. \cite{DSA05} claimed a very strong suppression of the Schiff moments of  by the RPA corrections. However,  Ref. \cite{HgCalc} demonstrated that in the case of $^{199}$Hg a more complete  Hartree-Fock-Bogoliubov calculation does not show such strong suppression as in the RPA method. In Ref. \cite{FKS1986} the ratio  $ S(^{1129}{\rm Xe})/S(^{199}{\rm Hg})=-1.25$. We use this result and Eq. (\ref{HgPbschiff}) for the $S(^{199}{\rm Hg})$ to estimate the $^{129}$Xe Schiff moment:
  \begin{align}\label{Xeschiff}
 S(^{129}{\rm Xe},g)  &\approx  ( - 0.03 g {\bar g}_0  + 0.01  g {\bar g}_1 - 0.04 g {\bar g}_2)\, e\cdot \textrm{fm}^3,\\
  S(^{129}{\rm Xe},{\bar \theta})  &\approx - 0.007  \,{\bar \theta} \, e\cdot \textrm{fm}^3 \ ,\\
  S(^{129}{\rm Xe},{\tilde d}) &\approx - 6 {\tilde d}_d\, e\cdot \textrm{fm}^2 \, .
 \end{align}
 We also present contributions of the neutron and proton EDM to the $^{129}$Xe Schiff moment  from Ref. \cite{DFS85}:
 \begin{equation} \label{dXe}
 S(^{129}{\rm Xe},d) \approx ( 0.63 d_n +0.13 d_p) \, \textrm{fm}^2
\end{equation} 
 

\begin{table}
\caption{\label{t:Schiff} Collective Schiff moments for deformed nuclei containing  opposite parity level with  the same spin close to the ground state. To obtain the values of the Schiff moments one should use   Eqs. (\ref{Sg},\ref{Stheta},\ref{Sd}) with the scaling factor $K_S$ presented in this table.
 For $^{225}$Ra we have  $K_S=1$ by definition -see Eq.~(\ref{Raschiff}). For nuclei $^{155}$Gd, $^{153}$Sm, $^{165}$Er not presented in this table we may expect $K_S   \sim 0.01 - 0.1$.}
\begin{ruledtabular}
\begin{tabular}{ccc}
\multicolumn{1}{c}{$Z$}&
\multicolumn{1}{c}{Isotope}&
\multicolumn{1}{c}{$K_S$}\\
\hline
63 & $^{153}$Eu & 3.7  \\
66 & $^{161}$Dy & $\lesssim 4$\\
66 & $^{163}$Dy & $\lesssim 0.4$  \\
86 & $^{223}$Rn  & 3  \\
87 & $^{221}$Fr & 0.14 \\
87 & $^{223}$Fr & 1.6 \\
89 & $^{225}$Ac  & 3  \\
89 & $^{227}$Ac  & 6 \\
90 & $^{229}$Th &  $ \lesssim  2$ \\
91 & $^{229}$Pa\tablenotemark[1]  & 40  \\
92 & $^{233}$U  & $\lesssim  2$  \\
92 & $^{235}$U  &  $ \lesssim  3$  \\
93 & $^{237}$Np  & 4 \\
94 & $^{239}$Pu  & $\lesssim 0.12$ \\
\end{tabular}
\end{ruledtabular}
\tablenotetext[1]{ Estimate for $^{229}$Pa  is presented assuming that the existence of a very close nuclear doublet level will be confirmed.}
\end{table}

\begin{table}
\caption{\label{t:SS-Schiff}Schiff moments 
in Xe, Hg, Tl and Pb nuclei.
$S(g) = (a g {\bar g}_0  + b  g {\bar g}_1 c g {\bar g}_2) \, [e \cdot {\rm fm}^3]$, 
$S(\bar \theta)= K_t \bar \theta \, [e \cdot {\rm fm}^3]$, $S({\tilde d}_d)=K_d {\tilde d}_d+K_d {\tilde d}_u \, [e \cdot {\rm fm}^2]$. Many-body corrections are expected to reduce the values of the Schiff moments of Tl nuclei presented in this table.}
\begin{ruledtabular}
\begin{tabular}{ldddddd}
\multicolumn{1}{c}{Isotope}&
\multicolumn{3}{c}{$S(g)$}&
\multicolumn{1}{c}{$S(\bar \theta)$}&
\multicolumn{1}{c}{$S({\tilde d}_d)$}\\
&\multicolumn{1}{c}{$a$}&
\multicolumn{1}{c}{$b$}&
\multicolumn{1}{c}{$c$}&
\multicolumn{1}{c}{$K_t$}&
\multicolumn{1}{c}{$K_d$}&
\multicolumn{1}{c}{$K_u$}\\
\hline
$^{129}$Xe &  -0.03 &   0.01 & -0.04  & -0.007 & -6 &  0\\
$^{203}$Tl,$^{205}$Tl &  0.13 &   -0.004 & -0.27  & 0.027 & 12 & 9 \\
$^{207}$Pb,$^{199}$Hg &  0.023 &   -0.007 & 0.029  & 0.005 & 5 & 0 \\
\end{tabular}
\end{ruledtabular}
\end{table}

\section{Atomic EDM calculations}

We have done calculations of the atomic electric dipole moments for few atoms in our earlier works~\cite{CI+MBPT,DFGK02,DFG07,DFP09,Kozlov}.
The EDM of an atom induced by $CP$-odd Hamiltonian $H_{\rm CP}$ is given by
\begin{equation}\label{eq:EDM}
\mathbf{d}_{\rm at} = 2\sum_n\frac{\langle 0|H_{\rm CP}|n\rangle\langle n|\mathbf{D}|0\rangle}{E_0-E_n},
\end{equation}
where $\mathbf{D} = \sum_i \mathbf{d}_i = -|e|\sum_i \mathbf{r_i}$ is the electric dipole operator with summation over all atomic electrons,
$|0\rangle$ is atomic ground state, summation in (\ref{eq:EDM}) goes over complete set of sates $|n\rangle$, $E_n$ are the energies of atomic states.

In this work we focus on atoms with enhanced nuclear  Schiff moments. Therefore, we consider only one type of the $CP$-odd Hamiltonian, the Hamiltonian of the interaction of atomic electrons with nuclear Schiff moment ($H_{\rm CP} \equiv H_{\rm SM}$). Taking into account the finite nuclear size effect, this Hamiltonian has the following form \cite{FG}:
\begin{equation}\label{eq:SM}
H_{\rm SM} =\sum_i h^{\rm SM}_i = -\sum_i \frac{3\mathbf{S\cdot r}_i}{B}\rho(r),
\end{equation}
where $B=\int(\rho(r)r^4dr$, $\rho$ is nuclear density, $\mathbf{S}$ is the vector of nuclear Schiff moment.
If $H_{\rm SM}$ in (\ref{eq:EDM}) is replaced by the electric dipole operator $\mathbf{D}$, the expression gives dipole static polarisability of the atom. This gives us  a good test of the accuracy of our  calculations.

The ways to reduce (\ref{eq:EDM}) to expressions containing only single-electron integrals depends on electron structure of the atom. Below we consider particular cases which cover almost all atoms in the periodic table. Note that the accuracy of the nuclear calculations of the Schiff moment  is never better than factor of 2 (see e.g. \cite{ERK}). Therefore,   30\% accuracy of the atomic calculations is sufficient. 

\subsection{Closed-shell atoms}

For closed-shell atoms, such as Xe, Rn, Fr$^+$, Ac$^{3+}$ and Th$^{4+}$, it is sufficient to use the so-called random-phase approximation (RPA).
The RPA equations present a linear response of the Hartree-Fock atomic states to a perturbation by an external field. They can be written in a form
\begin{equation}\label{eq:RPA}
(H_0 -\epsilon_c)\delta \psi_c = -(F+\delta V^F)\psi_c.
\end{equation}
Here $H_0$ is the relativistic Hartree-Fock (HF) Hamiltonian, $\psi_c$ is the HF electron state in the core, $\delta \psi_c$ is the correction to the HF state in the core induced by the external field $F$, $\delta V^F$ is the correction to the self-consistent HF
potential due to the corrections to all core states. Index $c$ numerates states in the core. Operator of the external field in our case is either  $H_{\rm SM}$ or $\mathbf{D}$ operators ($F=h^{\rm SM}$ or $F=d$).
RPA equations (\ref{eq:RPA}) are solved self-consistently for all states in the core.  The EDM of the closed-shell atom is given then by either of expressions
\begin{equation}\label{eq:EDMr}
\mathbf{d}_{\rm at} = \frac{2}{3}\sum_c \langle \psi_c | \mathbf{d}|\delta \psi_c^{\rm SM} \rangle,
\end{equation}
or
\begin{equation}\label{eq:EDMsm}
\mathbf{d}_{\rm at} = \frac{2}{3}\sum_c \langle \psi_c | h^{\rm SM}|\delta \psi_c^{\rm D} \rangle,
\end{equation}
where summation goes over all core states, $\delta \psi_c^{\rm SM}$ is the solution of the RPA equations with the Schiff moment operator $h^{\rm SM}$ and  $\delta \psi_c^{\rm D}$ is the solution of the RPA equations with the electric dipole operator $d$. Expressions (\ref{eq:EDMr}) and (\ref{eq:EDMsm}) give identical answers. 
Note that  replacing 
$\delta \psi_c^{\rm SM}$ in (\ref{eq:EDMr}) by $\delta \psi_c^{\rm D}$ or $h^{\rm SM}$ in (\ref{eq:EDMsm}) by $d$ gives the dipole static polarisability of the closed-shell atom (compare with  Eq. (\ref{eq:EDM})).
 
Expressions (\ref{eq:EDMr}) and (\ref{eq:EDMsm}) do not include inter-electron correlations beyond core polarisation which is treated by the RPA equations. They give accurate results for noble-gas atoms such as Xe and Rn where such correlations are small. Formally, they could  be used to calculate  EDM of such atoms as  Yb, Hg and Ra which have upper closed $s^2$ shell. However, the results would be less accurate. This is because in these atoms correlations between external electrons ($6s$ or $7s$) and between external electrons and electrons in the core are not small. Inclusion of  these correlations is considered in the next section.

\subsection{Atoms with few valence electrons}

In this section we consider atoms like Hg, Yb, Ra, Pb. Although some of these atoms can be considered as closed-shell atoms, the correlations affecting external electrons are important and should be taken into account for accurate results. This can be illustrated by calculation of atomic polarisability. The RPA value for the static polarisability of Hg is 44 a.u.~\cite{Dzuba16}. inclusion of correlations for two external electrons reduces the value to 34~a.u. brining it to excellent agreement with experiment (see Table~\ref{t:all}). The trend is confirmed by other calculations
(see, e.g.~\cite{Timo,DFGK02}).
The standard approach is to divide all electrons into two groups, most electrons go into the closed-shell core while few remaining electrons are treated as valence ones. Usual method of the calculations, CI+MBPT,  is the configuration interaction (CI) technique enhanced by the many-body perturbation theory (MBPT) approach for the more accurate calculation of the CI matrix elements.

EDM is calculated as a sum of the core and valence contributions. Core contribution is calculated as in previous section with one important amendment which is often called the core-valence contribution. Summation over excited states must not include states occupied by valence electrons due to Pauli exclusion principle. Since we do not perform the summation directly but rather solve the RPA equations, we take into account Pauli principle by imposing orthogonality condition on the RPA corrections to the core wave functions (see Eq.~(\ref{eq:RPA}))
\begin{equation} \label{eq:ort}
\delta \psi^{\prime}_c = \delta \psi_c - \langle \delta \psi_c| v \rangle \langle v|,
\end{equation}
where $v$ is the state occupied by an external electron.  

For valence electrons we use the CI+MBPT method~\cite{CI+MBPT}. 
The effective Hamiltonian has one- and two-electron terms:
\begin{eqnarray}\label{eq:CI}
H^{\rm CI} = \sum_i \left( H_0(r_i) + \Sigma_1(r_i) \right) + \\
\sum_{i<j} \left( \frac{e^2}{r_{ij}} + \Sigma_2(r_i,r_j) \right). \nonumber
\end{eqnarray}
Here $H_0$ is the relativistic HF Hamiltonian, $\Sigma_1$ and  $\Sigma_2$ are the operators responsible for the core-valence correlations~\cite{CI+MBPT}.
Summation in (\ref{eq:CI}) goes over valence electrons. We use the B-spline technique~\cite{B-spline} to construct single-electron orbitals in valence space. The wave functions for the valence electrons are found as an eigenstates of the CI Hamiltonian~(\ref{eq:CI}). They are used to calculate valence contribution to the EDM using formula (\ref{eq:EDM}). 

\subsection{Atoms with open $f$-shell}

These atoms are the main subject of present work. This is because these atoms have isotopes in which Schiff moment is strongly enhanced due to nuclear  octupole deformation. These include rare-earth and actinide atoms  $^{153}$Eu, $^{233}$U, $^{237}$Np, etc.~\cite{FF19}.
Accurate calculations for such atoms are difficult due to large number of electrons in open shells. One possible approach is to include all these electrons into valence space (see, e.g.~\cite{Timo2,CIPT}). This works well for energy levels and transition amplitudes~\cite{Yb-clock,SHE1,SHE2,SHE3,SHE4}. However, calculation of atomic characteristics, such as polarisabilities, EDM, etc., for which summation over complete set of intermediate states is needed, is problematic. This is because the methods are constructed for low energy states and not appropriate for high energy states needed for completeness.  
Therefore,  we use an alternative approach developed in Ref.~\cite{Kozlov} for polarisabilities of open-shell atoms. It was shown in \cite{Kozlov} that calculations can be performed as for two or three valence electron systems by attributing $f$-electrons to the core. Calculations for the core containing open $f$-shell are done as for closed-shell  system but with the use of the fractional occupation numbers for the $f$-shell. Calculations for the valence electrons are done with the use of the CI+MBPT method as in previous section. Note that atomic polarisability is given by the square of the electric dipole operator matrix element. In fact, the approached developed in \cite{Kozlov} is valid for any pair of vector operators, such as e.g. Schiff moment operator and electric dipole operator as in present work.

\subsection{Results of atomic calculations}

\begin{table*}
\caption{\label{t:all}Static scalar polarisabilities (in a.u.) and EMDs due to Schiff moment (in units of $10^{-17} [e \cdot {\rm cm}] S/[e \cdot {\rm fm}^3] $) for selected atoms and ions.
}
\begin{ruledtabular}
\begin{tabular}{ll ll rrrr rrrl}
\multicolumn{1}{c}{$Z$}&
\multicolumn{1}{c}{Atom}&
\multicolumn{2}{c}{Configuration}&
\multicolumn{4}{c}{Polarisability}&
\multicolumn{3}{c}{EDM} \\
&&&&\multicolumn{3}{c}{This work}&
\multicolumn{1}{c}{Other\tablenotemark[1]}&
\multicolumn{3}{c}{This work}&
\multicolumn{1}{c}{Other}\\
&&\multicolumn{1}{c}{Core}&
\multicolumn{1}{c}{Valence}&
\multicolumn{1}{c}{Core}&
\multicolumn{1}{c}{Valence}&
\multicolumn{1}{c}{Total}&&
\multicolumn{1}{c}{Core}&
\multicolumn{1}{c}{Valence}&
\multicolumn{1}{c}{Total}& \\
\hline
54 & Xe & [Pd]$5s^25p^6$ & none & 27 & none & 27 & 27.32(20) & 0.38 & none & 0.38 & 0.378~\cite{Latha}\\
63 & Eu & [Ba]$4f^{7}$ & $6s^2$ & 10 & 178 & 188 & 184(20) & 0.22 & -1.85 & -1.63 &\\
63 & Eu$^{3+}$ & [Ba]$4f^{6}$ & none & 7 & none & 7 &  & 0.33 & none & 0.33 &\\
64 & Gd & [Ba]$4f^{7}$ & $6s^25d$ & 9 & 150 & 159 & 158(20) & 0.24 & -2.36 & -2.22 &\\
66 & Dy & [Ba]$4f^{10}$ & $6s^2$ & 8 & 156 & 164 & 163(15) & 0.27 & -2.50 & -2.23 &\\
70 & Yb & [Ba]$4f^{14}$ & $6s^2$ & 6 & 141 & 147 & 139(6) & 0.37 & -2.26 & -1.88 &-1.903~\cite{Latha}; -1.9~\cite{DFG07}; -2.12~\cite{DFP09}\\
80 & Hg & [Yb]$5d^{10}$ & $6s^2$ & 7 & 27 & 34 & 33.91(34) & 0.37 & -2.87 & -2.50 &-2.63~\cite{DFP09}; -5.07~\cite{Latha1}\\
     &       &  &  &  &  &  &  &  &  &  &-2.914~\cite{Latha2}; -2.3~\cite{Skripnikov}\\
81 & Tl$^+$ & [Yb]$5d^{10}$ & $6s^2$ & 5 & 14 & 19 &  & 0.40 & -3.19 & -2.79 &\\
82 & Pb & [Hg] & $6s^26p^2$ & 3 & 43 & 46 & 47(3) & 0.39 & 0.15 & 0.54 & 0.766~\cite{Latha2}\\
82 & Pb$^{++}$ & [Hg] & $6s^2$ & 3 & 10 & 13 &  & 0.39 & -3.38 & -2.99 & -3.08~\cite{Latha2}\\
86 & Rn & [Hg]$6p^6$ & none & 35 & none & 35 & 35(2) & 3.3 & none & 3.3 & 3.3~\cite{DFP09}\\
87 & Fr$^+$ & [Rn] & $7s$ & 20 & none & 20 &  & 2.87 & none & 2.87 & \\
88 & Ra & [Rn] & $7s^2$ & 12 & 238 & 250 & 246(4) & 1.75 & -10.0 & -8.25 &-8.093~\cite{Latha}; -8.5~\cite{DFGK02}; -8.8\cite{DFP09}\\
89 & Ac & [Rn] & $7s^26d$ & 9 & 186 & 195 & 203(12) & 1.5 & -11.6 & -10.1 &\\
     & Ac$^+$ & [Rn] & $7s^2$ & 9 & 112 & 121 &  & 1.5 & -11.3 & -9.8 &\\
     & Ac$^{3+}$ & [Rn] & none & 10 & none & 10 &  & 2.5 & none & 2.5 &\\
90 & Th & [Rn] & $7s^26d^2$ & 6.7 &  &  & 217(54) & 1.4 & -10.6 & -9.2 &\\
     & Th$^{2+}$ & [Rn] & $7s^2$ & 6.7 & 45.7 & 52.4 &  & 1.4 & -8.33 & -6.93 &\\
     & Th$^{4+}$ & [Rn] & none & 7.8 & none & 7.8 &  & 2.4 & none & 2.4 &\\
91 & Pa & [Rn]$5f^{2}$ & $7s^26d$ & 20 & 150 & 170 & 154(20) & 0.5 & -11.9 & -11.4 &\\
92 &  U & [Rn]$5f^{3}$ & $7s^26d$ & 17 & 148 & 165 & 129(17) & 0.7 & -12.8 & -12.1 &\\
93 & Np & [Rn]$5f^{4}$ & $7s^26d$ & 14 & 146 & 160 & 151(20) & 0.9 & -8.4 & -7.5 &\\
94 & Pu & [Rn]$5f^{6}$ & $7s^2$ & 21 & 123 & 144 & 132(20) & 2.3 & -11.5 &  -9.2 &-10.9~\cite{DFGK02}\\
\end{tabular}
\end{ruledtabular}
\tablenotetext[1]{Recommended values from Ref.~\cite{Peter}}
\end{table*}

The results of atomic calculations are presented in Table~\ref{t:all}. In addition to atomic EDM calculated for the first time  we included atoms for which previous calculations are available. This is done to illustrate the accuracy of the calculations. Therefore, the list of previous calculations is far from being complete. 
In most cases we have included most recent or most accurate calculations. The table also includes the data on the calculated and experimental polarizabilities of neutral atoms. Studying these data provides another way of judging on the accuracy of the calculations. Note however, that atomic EDMs are sensitive to the wave functions in the vicinity of the nucleus while polarizabilities are not sensitive to it. Therefore, having good accuracy for polarizabilities is a necessary but not sufficient condition of having good accuracy for EDMs. For this reason we have studied the sensitivity of the calculated EDMs to the variation of the single-electron basis. We found, that more accurate results can be obtained if most important valence states ($6s$, $6p$  for atoms from Eu to Pb, $7s$, $7p$  for atoms from Ra to Pu) are calculated by the relativistic Hartree-Fock computer code without using B-splines.
For higher virtual states we use linear combinations of B-splines which are eigenstates of the relativistic HF Hamiltonian. The same results can be achieved if the number of B-splines is increased significantly and B-splined states are used everywhere, including the $6s$, $6p$, $7s$ and $7p$ states.  This modification of the basis is the main reason for some difference ($\sim 10\%$) with our earlier results of Refs.~\cite{DFGK02,DFG07,DFP09}. Taking this difference as an estimation of the accuracy of the calculations we see that the uncertainty is less than 30\%. This is a satisfactory accuracy since the accuracy of the interpretation of the EDM measurements in terms of fundamental parameters of the $TP$-odd interactions is limited by nuclear physics, where the accuracy is significantly lower. 

 Comparing our results with the results of other groups is another way to estimate uncertainties. The most striking difference is about two times disagreement of our result for Hg with the result of Latha et al Ref. ~\cite{Latha1}. Note however, that the later result of  Latha with different co-authors is about 1.7 times smaller~\cite{Latha2}. This new value as well as the results of other Hg EDM calculations agree with our results. Results of  Ref.~\cite{Latha2} for Pb and Pb$^{++}$ are also in agreement with our results. For other atoms the differences with results of other groups are within declared uncertainty.

Table~\ref{t:all} also includes some ions. Corresponding atoms can be found in molecules or crystals used in experimental search for CP-odd nuclear forces. The effect of the CP-violating forces in such systems can be reduced to the EDM of the ion of the heavy element. For example, the Ac$^+$ ion is a part of the AcF molecule suggested for the EDM measurements~\cite{FF19},  the Pb$^{2+}$ ion is a part of the PbO molecule and PbTiO$_3$ solid used in CASPEr experiment ~\cite{Casper}, the Th$^{2+}$ ion is a part of the ThO molecule used in the experiment Ref. ~\cite{ThO}. Note that the ground state of the isolated Th$^{2+}$ ion is [Rn]$5f6d \ ^3$H$^{\rm o}_4$. However, in the ThO molecule it is [Rn]$7s^2 \ ^1$S$_0$~\cite{ThO}.

To help in estimation of the sensitivity of different experiments, in the Table~\ref{t:theta} we presented values of the Schiff moments and atomic EDM for a number of most interesting atoms and ions in terms of the QCD $\theta$-term constant  ${\bar \theta}$. Dependence on other CP-violating parameters for all atoms and nuclear isotopes considered in the present paper may be found by the multiplication  of the results of the atomic calculations from Table~\ref{t:all} by the values of the Schiff moments presented in the section \ref{Schiff}.

\begin{table}
\caption{\label{t:theta}Schiff moments ($S$) and EDMs ($d_A$) of some atoms in terms of the QCD $\theta$-term constant  ${\bar \theta}$.  Remind the reader that the current experimental limit is $|{\bar \theta}|<10^{-10}$.}
\begin{ruledtabular}
\begin{tabular}{ccccc}
\multicolumn{1}{c}{$Z$}&
\multicolumn{1}{c}{Atom}&
\multicolumn{1}{c}{$S$}&
\multicolumn{2}{c}{$d_A [e \cdot {\rm cm}]$}\\   
&&\multicolumn{1}{c}{${[e \cdot \ {\rm fm}^3 {\bar \theta}]}$}&
\multicolumn{1}{c}{$10^{-17}S [e \cdot \ {\rm fm}^3]  $}&
\multicolumn{1}{c}{$10^{-17} {\bar \theta} $} \\
\hline
63 & $^{153}$Eu              & -3.7 & -1.63 & 6 \\
63 & $^{153}$Eu$^{3+}$  & -3.7 & 0.33 & -1.2 \\
66 & $^{161}$Dy              & $\lesssim$ 4 & -2.23 & $\lesssim$9 \\
80 & $^{199}$Hg              & 0.005 & -2.50 & -0.013 \\
81 & $^{205,203}$Tl$^+$ & 0.02 & -2.79 & -0.06 \\
82 & $^{207}$Pb$^{2+}$  & 0.005 & -2.99 & - 0.015 \\
86 & $^{223}$Rn              & -3 & 3.3 & -10 \\
87 & $^{223}$Fr$^+$     & -1.6 & 2.87 & -4.6 \\
88 & $^{225}$Ra            & -1 & -8.25 & 8 \\
89 & $^{227}$Ac            & -6 & -10.1 & 60 \\
89 & $^{227}$Ac$^+$     & -6 & -9.8 & 60 \\
90 & $^{229}$Th$^{2+}$  &$\lesssim$ 2   &-6.93   &  $\lesssim$ 14\\
91 & $^{229}$Pa\tablenotemark[1]           & -40 & -11.4 & 460 \\
92 & $^{233}$U             &$\lesssim$ 2 & -12.1 & $\lesssim$ 20 \\
93 & $^{237}$Np           & -4 & -7.5 & 30 \\
94 & $^{239}$Pu           &$\lesssim$  0.1 & -9.2 &$\lesssim$ 1 \\
\end{tabular}
\end{ruledtabular}
\tablenotetext[1]{ Estimates for $^{229}$Pa  are presented assuming that the existence of a very close nuclear doublet level will be confirmed.}
\end{table}

\section{Molecular T,P-violating spin-axis interaction constants}

Molecules have very close opposite parity rotational levels or rotational doublets. Polar molecules also have large intrinsic electric dipole moments and may be nearly completely polarised by an external electric field. In this case an internal molecular field (which for polarised molecule is directed along the external electric field) exceeds an external field by many orders of magnitude and dramatically increases sensitivity to the T,P-violating interactions \cite{Sandars,SushkovFlambaum}. For example, experiment with ThO molecule \cite{TheEDM} gives a limit on electron EDM which is two orders of magnitude better then the limit from the atomic EDM measurement. Sandars  \cite{Sandars} suggested to use this molecular property to measure T,P- violating interaction  of the nuclear spin with molecular axis. Similar to the atomic EDM, this interactions is induced by the electrostatic  interaction of the nuclear Schiff moment with molecular  electrons  which are able to reach heavy nucleus. The effect increases faster than $Z^2$ \cite{SFK} so  actinide   molecules which have enhanced Schiff moment may have an advantage.

    The interaction constant $W_S$ for the effective T,P-violating interaction in molecules is defined by the following expression:
 \begin{equation}\label{WSdefinition}   
   W_{T,P}=W_S\frac{S}{J} {\bf J \cdot n}\,,  
 \end{equation} 
where ${\bf J}$ is the nuclear spin, ${\bf n}$ is  the  unit vector along the molecular axis in linear molecules \footnote{Some older molecular calculations presented constant $X=W_S/6$.}. 

Molecule AcF has valence electron structure similar to TlF. Indeed assuming that upper valence electron goes to F atom,   Ac$^+$ and Tl$^+$ ions have valence $7s^2$ and $6s^2$ correspondingly. The ratio of atomic EDM for Ac$^+$ and Tl$^+$ is 3.5 (see Table  \ref{t:all}) so we have 
\begin{equation}\label{WsAc}   
  W_S( {\rm AcF})=3.5 W_S( {\rm TlF}) \sim 100000
 \end{equation} 
in atomic units (here a.u.=$e/a_B^4$). Here we used $W_S( {\rm TlF})$=45810 from Ref. \cite{Petrov} which includes many-body effects reducing the result by 22 \%.
Previous calculations for TlF have been published in Refs. \cite{Hinds,Parpia,Quiney}.
Note, however, that our previous experience with the extrapolation from TlF to RaO (see Refs. \cite{RaO,RaOTitov} ) indicates that  $W_S( {\rm AcF})$ in Eq. (\ref{WsAc}) may be strongly overestimated since Tl and Ac atoms chemical properties are very different.    

Substitution of the Schiff moment from Table \ref{t:Schiff}  to the energy shift $W_{T,P}=W_S\frac{S}{J} {\bf J \cdot n}$ gives for the fully polarised molecule the energy difference between the $J_z=J$ and $J_z=-J$ states:
 \begin{equation}\label{WActheta} 
 2 W_S( {\rm AcF})S (^{227}{\rm Ac}) \sim  10^8  {\bar \theta} \, h\, \textrm{Hz},
 \end{equation}    
  where $h$ is the Plank constant.  This is 1000 times larger than the energy shift for TlF: 
   \begin{equation}\label{WTltheta} 
  2 W_S( {\rm TlF})  S (^{205}{\rm Tl})=1.0 \cdot 10^5  {\bar \theta} \, h\, \textrm{Hz}.
  \end{equation} 
  Using the Hg atom EDM  measurements  Ref. \cite{HgEDM} and Eqs. (\ref{HgPbschiff}) and  (\ref{dHg2})  we can extract the limits $|{\bar \theta}|< 5\cdot 10^{-11}$, $|d_n| <10^{-26} e \cdot$ cm and $|d_p| < 5 \cdot 10^{-25} e \cdot$ cm. The neutron EDM measurements give limit $| {\bar \theta} | <  10^{-10}$. With this limit  the maximal shift in AcF is $\sim  \cdot 10^{-3}$ Hz. 
    The measured shift in the 1991 TlF experiment \cite{TlFexperiment} was $(-1.3 \pm 2.2) \cdot 10^{-4}$ Hz, i.e. such accuracy is already sufficient.  
  It is expected that new generation of molecular experiments will improve this accuracy by several orders of magnitude \cite{NewTlF}.  Therefore, we may expect a very significant  
 improvement of  the current limit  $| {\bar \theta} | < 10^{-10}$ and also improvement of the limits on other fundamental parameters of the CP-violation theories such as the $\pi NN$ interaction constants ${\bar g}$ and the quark chromo-EDMs ${\tilde d}$.
 
Other interesting examples include  molecules AcN and AcO$^+$ which have electronic structure similar to RaO (naively, they may be considered as ion molecules Ac$^{3+}$N$^{3-}$, Ac$^{3+}$O$^{2-}$ and Ra$^{2+}$O$^{2-}$ where corresponding atomic ions have the same electron states). The ratio of Ac$^+$ EDM  to  Ra EDM is 9.8/8.25=1.18 (see Table \ref{t:all}), so we may use the results  of the RaO calculation  $W_S( {\rm RaO})=45192$ from  Ref. \cite{RaOTitov} to obtain  reliable estimates for  AcN and AcO$^+$:
\begin{equation}\label{WsAcN}   
W_S( {\rm AcN}) \approx  W_S( {\rm AcO}^+)  \approx 53000.
 \end{equation}

 If we assume that in  PbO molecule two electrons go from Pb to O, it has  electronic  structure similar to TfF.  The ratio of atomic EDM for Pb$^{2+}$ and Tl$^+$ is 1.06 (see Table  \ref{t:all}) so we have 
\begin{equation}\label{WsPb}   
  W_S( {\rm PbO})=1.06 W_S( {\rm TlF})=49,000\,.
 \end{equation} 
A more accurate direct calculation in Ref. \cite{PbO} gives 47250. Ref.  \cite{PbO} also gives value of $W_s$ in the solid PbTiO$_3$  used in the CASPEr experiment ~\cite{Casper}:   
\begin{equation}\label{WsPbTiO}   
  W_S( {\rm PbTiO}_3)=30270 .
 \end{equation} 
The Schiff moment of $^{207}$Pb is  slightly smaller than that of $^{205}$Tl since $^{207}$Pb has valence neutron. 

Ground state of Eu$3+$ ion has zero electron angular momentum. Possibly, some Eu molecules may have ground or metastable state with the zero angular momentum too. Unfortunately, we have not found  a specific example yet.  Possibly such  state exists in the molecule  EuN where 3 electrons from Eu atom can make close shell on N atom. Another example is molecular ion EuO$^+$ which has electronic  structure similar to EuN.  The estimate for the energy shift is:   
\begin{equation}\label{WEutheta} 
 2 W_S( {\rm EuN})S (^{153}{\rm Eu}) \approx 2 W_S( {\rm EuN})S (^{153}{\rm Eu}) \sim  10^7  {\bar \theta} \, h\, \textrm{Hz}.
 \end{equation}  
 This is 100 times larger than the energy shift for TlF.   

Th atom has two extra $6d^2$ electrons in comparison with Ra atom. $6d^2$ electrons  give a very small contribution to atomic EDM, therefore the atomic calculation of Th EDM gives the  result close to that for Th$^{2+}$  and Ra EDM.  Th$^{2+}$ and Ra have similar electronic structure with filled $7s^2$ subshell and close values of atomic EDM in units of the Schiff moment  (see Table  \ref{t:all}).  Therefore,  $W_S$ for the ThO molecule is comparable to that of RaO molecule calculated in   Ref. \cite{RaOTitov}:
\begin{equation}\label{WSZ}   
W_S( {\rm ThO}) \sim W_S( {\rm RaO})=45192 
 \end{equation}   
 Another possibility may be  to use  the doublet in $^3\Delta_1$ metastable state of $^{229}$ThO (used  to improve the limit on electron EDM in Ref. \cite{ThO}) and the ground state  doublet  $^3\Delta_1$ in ThF$^+$. 
 
   D. DeMille and T. Fleig suggested recently to make cold molecule AgRa for EDM measurements from cold atoms cooled by laser \cite{FleigCanberra}.   Some atoms which we considered may be cooled, including Fr, Ac and Ra. There is a number of  atoms which may be cooled and  combined with Fr, Ac and Ra atoms to form a cold molecule. Corresponding T,P-odd interaction   constant $W_S$ is determined mainly by the heavy atoms  and may be estimated using the molecular calculations for TlF and RaO using the results for Fr$^+$, Ac$^+$, Ra$^{2+}$ and Tl$^{+}$  EDM.

  Finally, in the recent paper \cite{MOH+} it was suggested that linear molecules MOH, molecular ions MOH$^+$ (M is a heavy atom, e.g. Ra in the molecule RaOH$^+$ ) and symmetric top molecules (such as MCH$_3$ or MOCH$_3$) may be better systems than molecules MO since such polyatomic molecules have a doublet of close opposite parity energy levels in the bending mode and may be polarised by a weak electric field. The reduction of the  strength of necessary electric field simplifies the experiment and dramatically reduces systematic effects. These molecules may be cooled by a laser.
  
\section{Conclusion} 
In this paper we presented estimates for the nuclear Schiff moments, calculations of the atomic EDMs produced by these Schiff moments and estimates of the T,P-odd nuclear spin - molecular axis interaction constants for molecules containing these atoms. For the comparison and estimate of the accuracy we included atoms  where the  measurements as well as the  atomic and nuclear calculations have been done previously:  $^{199}$Hg, $^{129}$Xe, $^{225}$Ra.  

Then we presented results for  atoms and ions   where the  nuclear Schiff moments and atomic EDMs expressed in terms of the CP-violating $\pi$-meson - nucleon interaction constants ${\bar g}_0$, ${\bar g}_1$, ${\bar g}_2$, QCD parameter ${\bar \theta}$ and quark chromo-EDM were not available  including Pb, Tl$^+$ and  Pb$^{2+}$. These ions may be considered as parts of the TlF and PbO molecules and  PbTiO$_3$ solid used in the CASPEr experiment ~\cite{Casper} searching for the axion dark matter which induces Pb$^{2+}$ oscillating  EDM. 

The main part of the work is the calculations for stable or very long lifetime rare-earth and actinide atoms and ions  where the nuclear Schiff moments are enhanced up to three orders of magnitude:  Eu, Eu$^{3+}$, Gd, Dy, Ac, Ac$^+$,  Ac$^{3+}$,  Th, Th$^{2+}$, Th$^{4+}$, Pa, U, Np, Pu. In the molecules containing corresponding ions, e.g.  in $^{227}$AcF or $^{227}$AcN, the T,P-violating effects are up to three orders of magnitude larger than in TlF where the experiments have been performed earlier \cite{TlFexperiment} and are carried out now \cite{NewTlF}.

\section{Acknowledgements}
This work was supported by the Australian Research Council and the Gutenberg Fellowship. We are grateful to Hans Feldmeier and  Leonid Skripnikov for useful discussions.

\bibliographystyle{apsrev}

\end{document}